\documentclass{raa}

\usepackage[varg]{txfonts}

\usepackage{natbib}
\usepackage{graphicx}
\usepackage{epstopdf}
\usepackage{amssymb}
\usepackage{float}
\usepackage{longtable}

\bibpunct{(}{)}{;}{a}{}{,}

\begin{document}

\title{Orbital period correction and light curve modeling of the W-subtype shallow contact binary OW Leo}

   \volnopage{Vol.0 (200x) No.0, 000--000}      
   \setcounter{page}{1}          

\author{Xiao Zhou
      \inst{1,2,3}
   \and Shengbang Qian
      \inst{1,2,3,4}
}

\institute{
             Yunnan Observatories, Chinese Academy of Sciences (CAS), P.O. Box 110, 650216 Kunming, P. R. China {\it zhouxiaophy@ynao.ac.cn}\\
        \and
             Key Laboratory of the Structure and Evolution of Celestial Objects, Chinese Academy of Sciences, P. O. Box 110, 650216 Kunming, P. R. China\\
        \and
             Center for Astronomical Mega-Science, Chinese Academy of Sciences, 20A Datun Road, Chaoyang Dis-trict, Beijing, 100012, P. R. China\\
        \and
             University of the Chinese Academy of Sciences, Yuquan Road 19\#, Sijingshang Block, 100049 Beijing, P. R. China\\
}

\date{Received~~2017 month day; accepted~~2017~~month day}

\def\gsim{\mathrel{\raise.5ex\hbox{$>$}\mkern-14mu
                \lower0.6ex\hbox{$\sim$}}}

\def\lsim{\mathrel{\raise.3ex\hbox{$<$}\mkern-14mu
               \lower0.6ex\hbox{$\sim$}}}

\abstract{
Orbital period and multi-color light curves' investigation of OW Leo are presented for the first time. The orbital period of OW Leo is corrected from $P = 0.325545$ days to $P = 0.32554052$ days in our work, and the observational data from the All-Sky Automated Survey for SuperNovae (ASAS-SN) are used to test the newly determined orbital period. Then, the phased light curves are calculated with the new period and the Wilson-Devinney program is applied to model the light curves, which reveal that OW Leo is a W-subtype shallow contact binary system ($q = 3.05$, $f = 12.8\,\%$). The absolute physical parameters of the two component stars are estimated to be $M_{1} = 0.31(1)M_\odot$, $M_{2} = 0.95(3)M_\odot$, $R_{1} = 0.63(1)R_\odot$, $R_{2} = 1.04(1)R_\odot$, $L_{1} = 0.43(1)L_\odot$ and $L_{2} = 1.01(2)L_\odot$. The evolutionary status show that the more massive star is less evolved than the less massive star. OW Leo has very low metal abundance, which means its formation and evolution are hardly influenced by any additional component. It is formed from an initially detached binary systems through nuclear evolution and angular momentum loss via magnetic braking, and have passed a very long time of main sequence evolution.
}

\authorrunning{Zhou et al.}
\titlerunning{W-subtype shallow contact binary of OW Leo}

\maketitle


\section{Introduction}\label{intro}

Contact binaries, both inside and outside our Galaxy, have been proved to be powerful tools for studying a wide spectrum of astrophysical problems. They usually consist of F, G and K type stars and the component stars are embedded in a common envelope. Due to the strong interactions between the component stars, the formation and evolution scenario for contact binaries are entirely and totally different from those of single stars, and still remains to be an open issue. A few years ago, many contact binaries have been observed by photometric method. However, only very few targets have spectral information due to the low efficiency of spectroscopic observations. Owe to the Large Sky Area Multi-Object Fibre Spectroscopic Telescope (LAMOST) \citep{2012RAA....12.1197C,2012RAA....12..723Z}, hundreds of thousands of contact binaries are provided with stellar atmospheric parameters, which makes it a better time to investigate contact binaries \citep{2017RAA....17...87Q,2018ApJS..235....5Q}.

OW Leo, also named 	GSC 01443-00087 or NSVS 10352593, is a newly reported EW-type binary \citep{2010IBVS.5945....1D}. Its period was given as $P = 0.325545$ days in the International Variable Star Index (VSX) \citep{2006SASS...25...47W}. The binary system was observed by the LAMOST project for two times, which were on March 24, 2013 and February 6, 2018. The low resolution spectra gave out two sets of stellar atmospheric parameters, which were listed in Table \ref{lamost}. There is no radial velocity curve published for the target and its light curves haven't been analysed yet. In the present work, we will state our photometric observation on OW Leo in Section 2. Then, all available mid-eclipse times are gathered to investigate the orbital period changes of OW Leo in Section 3, and our observational light curves and light curve from All-Sky Automated Survey for SuperNovae (ASAS-SN) are modeled to get its physical parameters in Section 4. Finally, the formation and evolution theory of OW Leo will be discussed in Section 5.
\begin{table}[!ht]\small
\begin{center}
\caption{Stellar atmospheric parameters of OW Leo.}\label{lamost}
\begin{tabular}{ccccccc}\hline\hline
    T (K)         &    Log (g)     & [Fe/H] &       Date        \\\hline\hline
    5605          &    4.131       &   -0.734    &  March 24, 2013   \\
    5724          &    4.250       &   -0.606    &  February 6, 2018   \\
\hline\hline
\end{tabular}
\end{center}
\end{table}

\section{OBSERVATIONS AND DATA REDUCTION}

\begin{figure}[!ht]
\begin{center}
\includegraphics[width=12cm]{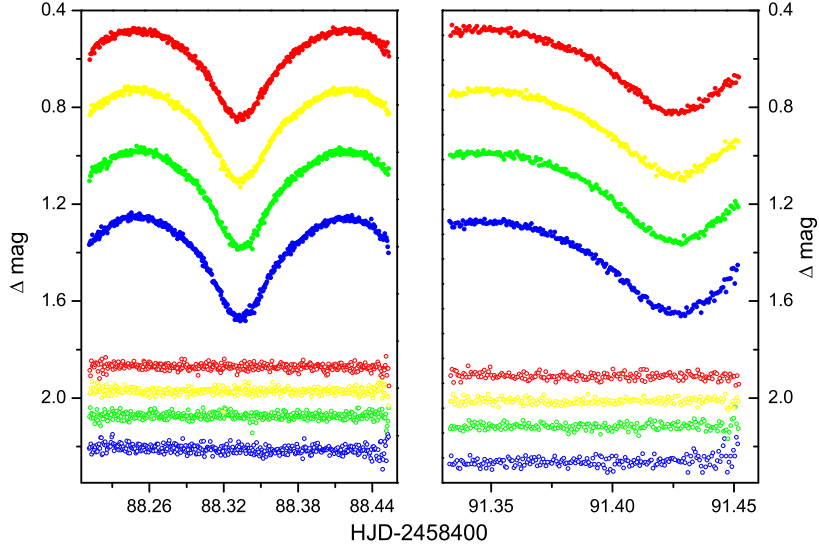}
\caption{The light curves obtained on January 5 are plotted in the left panel of Fig. 1 and the light curves obtained on January 8 are displayed in the right. The blue, green, yellow and red colors refer to light curves observed with $B$, $V$, $R_C$ and $I_C$ filters, respectively. The magnitude differences between the Variable star (OW Leo) and Comparison star (UCAC4 546-051376) are plotted with solid circles and the magnitude differences between the Comparison star (UCAC4 546-051376) and Check star (UCAC4 546-051360) are plotted with open circles.}\label{LC_Obs}
\end{center}
\end{figure}

The multi-color light curves of OW Leo were obtained on January 5 and 8, 2019 with the Xinglong 85cm optical telescope of National Astronomical Observatories, Chinese Academy of Sciences. The telescope was equipped with a 2K $\times$ 2K CCD Camera, and its field of view (FOV) was 32.8 square arc-minutes \citep{2009RAA.....9..349Z,2018MNRAS.478.3513Z}. Four filters ($B$, $V$, $R_C$ and $I_C$) from Johnson-Cousins' filters systems were chosen for the observations and the observational images were processed with the Image Reduction and Analysis Facility (IRAF) \citep{1986SPIE..627..733T}. UCAC4 546-051376 and UCAC4 546-051360 in the same field of view were chosen as the Comparison (C) star and Check (Ch) star since we used the differential photometry method to obtain the light variations of OW Leo. Their coordinates in J2000.0 epoch and magnitudes in $V$ band are listed in Table \ref{Coordinates}. Finally, we got the light variations of OW Leo and the observational light curves were plotted in Fig. \ref{LC_Obs}.

\begin{table}[!h]\small
\begin{center}
\caption{Coordinates and $V$ band magnitudes.}\label{Coordinates}
\begin{tabular}{cccc}\hline\hline
Target              &   $\alpha_{2000}$         &  $\delta_{2000}$          &  $V_{mag}$   \\ \hline\hline
OW Leo              &$11^{h}52^{m}13^{s}.59$    & $+18^\circ58'55''.0$      &  $12.84$     \\
UCAC4 546-051376    &$11^{h}52^{m}39^{s}.70$    & $+19^\circ03'00''.2$      &  $12.12$     \\
UCAC4 546-051360    &$11^{h}51^{m}50^{s}.00$    & $+19^\circ04'01''.1$      &  $13.98$     \\
\hline\hline
\end{tabular}
\end{center}
\end{table}

As shown in Fig. \ref{LC_Obs}, we got the primary minimum (Left of Fig. \ref{LC_Obs} ) and secondary minimum (Right of Fig. \ref{LC_Obs}) of the eclipsing binary system. Then, parabola fit were applied on the observed light minima curves and the mid-eclipse times were determined, which were listed in Table \ref{New_minimum}.

\begin{table}[!ht]\small
\begin{center}
\caption{New CCD times of mid-eclipse times.}\label{New_minimum}
\begin{tabular}{ccccccc}\hline\hline
    JD (Hel.)     &  Error (days)  &  p/s &           Filter        &       Date        \\\hline\hline
  2458488.3337    & $\pm0.0002$    &   p  &   $B$ $V$ $R_C$ $I_C$   &  January 5, 2019  \\
  2458491.4265    & $\pm0.0002$    &   s  &   $B$ $V$ $R_C$ $I_C$   &  January 8, 2019  \\
\hline\hline
\end{tabular}
\end{center}
\end{table}

\section{ORBITAL PERIOD INVESTIGATION }

Orbital period variations are quite common among close binary systems due to the possible angular momentum loss (ALM) from binary systems or mass transfer between the primary or secondary stars \citep{2018RAA....18...30Z,2018MNRAS.474.5199L}. Thus, the O - C method was widely used in correcting orbital period or calculating orbital period variations of close binaries \citep{2017AJ....154..260P,2019AdAst2019E...5H}. For analysis the orbital period of OW Leo, all available mid-eclipse times are gathered and listed Table \ref{Minimum}. The descriptions for each column of Table \ref{Minimum} are:

Column 1  - the observed mid-eclipse times in HJD;

Column 2  - primary (p) or secondary (s) light minima;

Column 3  - cycle numbers calculated from Equation \ref{Ephemeris};

Column 4  - the calculated $O - C$ values basing on Equation \ref{Ephemeris};

Column 5  - observational errors;

Column 6  - the references;

The $O - C$ values listed in Column 4 of Table \ref{Minimum} are calculated with the linear equation:

\begin{equation}
Min.I(HJD) = 2453175.509 + 0^{d}.325545\times{E}.\label{Ephemeris}
\end{equation}

\begin{table}[!h]
\caption{Mid-eclipse times and $O - C$ values of OW Leo.}\label{Minimum}
\begin{center}
\small
\begin{tabular}{ccllrrc}\hline\hline
  JD(Hel.)      &  p/s    &       Epoch       &       $O - C$      &   Error      & Ref.   \\
 (2400000+)     &         &                   &       (days)       &   (days)     &         \\\hline
53175.509 	    &   p	  &     0             &      0             &         	  &    1     \\
55269.7211 	    &   p	  &     6433          &      -0.0189       &   0.0005	  &    1     \\
55604.8643	    &   s	  &     7462.5        &      -0.0243       &   0.0006	  &    2     \\
55973.8650	    &   p	  &     8596          &      -0.0288       &   0.0007	  &    3    \\
56041.7390	    &   s	  &     8804.5        &      -0.0310       &   0.0021	  &    3     \\
58488.3337	    &   p	  &     16320         &      -0.0697       &   0.0002	  &    4     \\
58491.4265      &   s     &     16329.5       &      -0.0696       &   0.0001     &    4    \\\hline
\end{tabular}
\end{center}
\textbf
{\footnotesize Reference:} \footnotesize (1) \citet{2010IBVS.5945....1D}; (2) \citet{2011IBVS.5992....1D}; (3) \citet{2012IBVS.6029....1D}; (4) the present work.
\end{table}

\begin{figure}[!ht]
\begin{center}
\includegraphics[width=12cm]{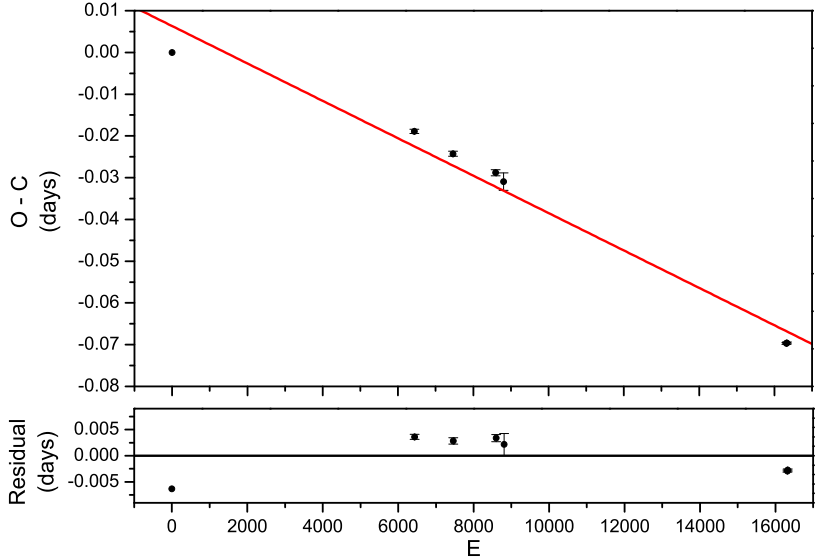}
\caption{The black dots in the upper panel refer to observational O - C values calculated from Equation \ref{Ephemeris} and the red line represents theoretical O - C curve based on Equation \ref{New_ephemeris}. The fitting residuals are shown in the lower panel of Fig. \ref{O-C}.}\label{O-C}
\end{center}
\end{figure}

As displayed in Fig. \ref{O-C}, linear fit is performed on the observed O - C curve, which means the orbital period of OW Leo should be corrected from 0.325545days to 0.32554052days. The new ephemeris is determined to be:
\begin{equation}\label{New_ephemeris}
\begin{array}{lll}
Min. I =  2453175.5153(\pm0.0002)+0^{d}.32554052(\pm0.00000002)\times{E}
\end{array}
\end{equation}

\section{LIGHT CURVE MODELING}

The modeling of OW Leo's light curves have been left out since the eclipsing binary system was discovered. We are going to modeling its light curves with the Wilson-Devinney program \citep{1971ApJ...166..605W}. The target have been observed by the LAMOST sky survey project for two times and the low resolution spectra show that OW Leo is an solar-type binary system. The averaged temperature of 5665K was assumed to the component star. The gravity-darkening coefficients are correspondingly assumed to be $g_{1} = g_{2} = 0.32$\citep{1967ZA.....65...89L}, and the bolometric albedo coefficients are assumed to be $A_{1} = A_{2} = 0.5$ \citep{1969AcA....19..245R}, and the limb darkening coefficients are chosen accordingly \citep{1993AJ....106.2096V}. The phased light curves are calculated from the linear ephemeris equation:

\begin{equation}
Min.I(HJD) = 2458488.3337+0^{d}.32554052\times{E}.\label{phase}
\end{equation}

\subsection{LIGHT CURVES FROM XINGLONG 85CM TELESCOPE}

\begin{figure}[!ht]
\begin{center}
\includegraphics[width=12cm]{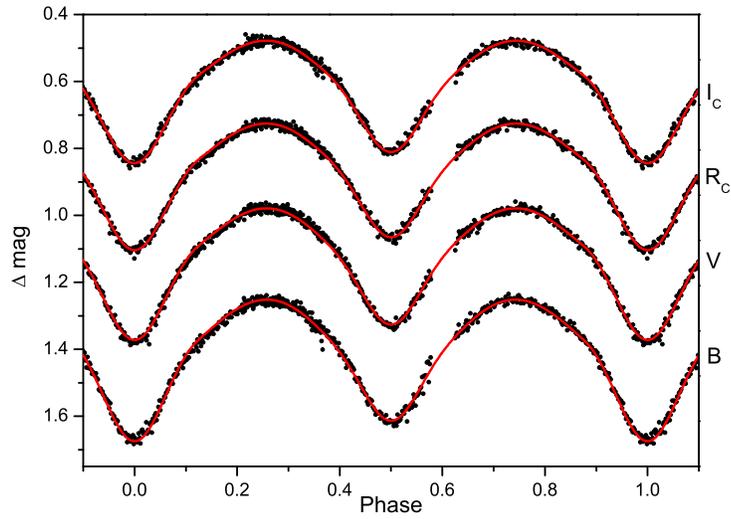}
\caption{The black dots are observational light curves obtained with the Xinglong 85cm optical telescope. The red lines are synthetic light curves from the Wilson-Devinney program.}\label{LC-BVRI}
\end{center}
\end{figure}

\begin{figure}[!ht]
\begin{center}
\includegraphics[width=10cm]{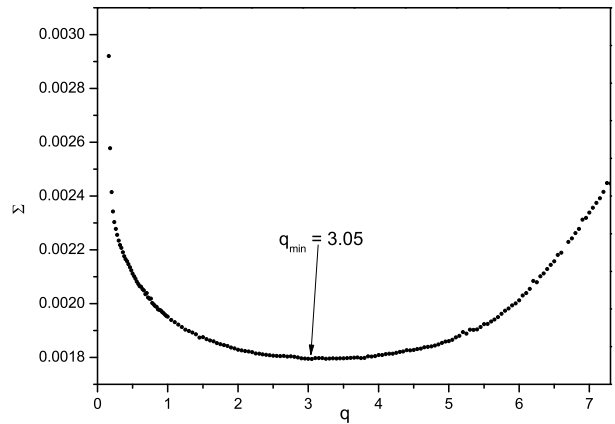}
\caption{The q-search diagram.}\label{q-search}
\end{center}
\end{figure}

As displayed in Fig. \ref{LC-BVRI}, the observed light curves are EW-type. Thus, Mode 3 for contact binary, Mode 4 and Mode 5 for semi-contact binary are chosen to modeling the observational light curves. Since there is no radial velocity curves published, the q-search method is used to get the initial mass ratio \citep{2018PASJ...70...87Z}. At first, the temperature of 5665K was assumed to the primary star ($T_{2}$), and the adjustable parameters are orbital inclination ($i$), modified dimensionless surface potential of the two component stars ($\Omega_{1}$, $\Omega_{2}$), mean surface temperature of the secondary star ($T_{2}$), bandpass luminosity of the two component stars ($L_{1}$, $L_{2}$). Then, we find that the light curves of OW Leo converged in Mode 3 and its mass ratio ($M_2/M_1$) should be larger than 1, which mean that OW Leo is a W-subtype contact binary system. Therefore, the temperature of 5665K is assumed to the secondary star ($T_{2}$). The corresponding q-search diagram are shown in Fig.\ref{q-search}. Then, the initial mass ratio is given to be $q = 3.05$ and set as a free parameter to run the Wilson-Devinney program again. Finally, the physical parameter of the binary system are determined, which are listed in Table \ref{WD_results}. The synthetic light curves are also plotted in Fig. \ref{LC-BVRI}. It should be pointed out that we also try to set third light ($l_{3}$) as a free parameter, but the light contribution from a tertiary component is not detected.

\begin{table}[!h]
\begin{center}
\caption{Photometric results of OW Leo}\label{WD_results}
\small
\begin{tabular}{lllllllll}
\hline\hline
Parameters                            &   Values                      \\\hline
$T_{2}(K)   $                         &  5665(assumed)                  \\
q ($M_2/M_1$ )                        &  3.05($\pm0.05$)                 \\
$i(^{\circ})$                         &  68.2($\pm0.1$)                    \\
$T_{1}(K)$                            &  5885($\pm5$)                       \\
$\Delta T(K)$                         &  220($\pm5$)                         \\
$L_{1}/(L_{1}+L_{2}$) ($B$)           &  0.3169($\pm0.0008$)                \\
$L_{1}/(L_{1}+L_{2}$) ($V$)           &  0.3041($\pm0.0006$)                 \\
$L_{1}/(L_{1}+L_{2}$) ($R_c$)         &  0.2981($\pm0.0005$)                \\
$L_{1}/(L_{1}+L_{2}$) ($I_c$)         &  0.2937($\pm0.0005$)                  \\
$r_{1}(pole)$                         &  0.2731($\pm0.0002$)                  \\
$r_{1}(side)$                         &  0.2852($\pm0.0003$)                   \\
$r_{1}(back)$                         &  0.3220($\pm0.0005$)                    \\
$r_{2}(pole)$                         &  0.4540($\pm0.0002$)                    \\
$r_{2}(side)$                         &  0.4883($\pm0.0003$)                    \\
$r_{2}(back)$                         &  0.5157($\pm0.0004$)                    \\
$\Omega_{1}=\Omega_{2}$               &  6.603($\pm0.003$)                     \\
$f$                                   &  $12.8\,\%$($\pm$0.5\,\%$$)      \\
$\Sigma{\omega(O-C)^2}$               &  0.0018                                 \\
\hline
\hline
\end{tabular}
\end{center}
\end{table}

\subsection{LIGHT CURVES FROM ASAS-SN}

\begin{figure}[!ht]
\begin{center}
\includegraphics[width=13cm]{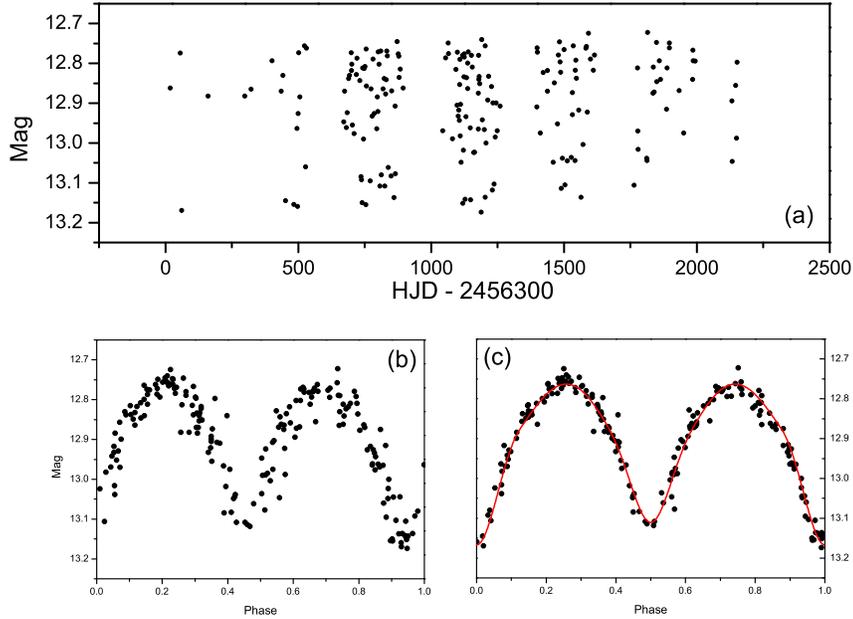}
\caption{Observational and synthetic light curves from ASAS-SN.}\label{ASAS-SN}
\end{center}
\end{figure}

OW Leo was also observed by the All-Sky Automated Survey for SuperNovae (ASAS-SN) \citep{2014ApJ...788...48S,2018MNRAS.477.3145J} from January 25, 2013 to November 28, 2018. The data from the ASAS-SN was plotted in Fig. \ref{ASAS-SN} (a). To modeling the data, we converted it to phased light curve. The light curve in Fig. \ref{ASAS-SN} (b) was calculated basing on the original orbital period $P = 0.325545$ days and the light curve in Fig. \ref{ASAS-SN} (c) was calculated with the corrected orbital period $P = 0.32554052$ days. It is obviously that the orbital period correction is necessary. Then, the phased light curve in Fig. \ref{ASAS-SN} (c) was also modeled with the Wilson-Devinney program, and the synthetic light curve was plotted with red line in Fig. \ref{ASAS-SN} (c).

\section{DISCUSSION AND CONCLUSION}

The orbital period and light curves of OW Leo are investigated for the first time, which reveal that its orbital period should be corrected from $P = 0.325545$ days to $P = 0.32554052$ days. According to the photometric results listed in Table \ref{WD_results}, we can conclude that OW Leo is a W-subtype shallow contact binary system with its mass ratio to be $q = 3.05(5)$ and fill-out factor to be $f = 12.8(5)\,\%$. The mass of the secondary star is estimated to be $ M_2 = 0.95(3) M_\odot$ basing on its effective temperature of $ T_2 = 5665K$ \citep{Cox2000}. Then, the other absolute parameters of the two component stars are calculated and listed in Table \ref{absolute}. And the orbital semi-major axis is calculated to be $a = 2.15(\pm0.02)R_\odot$ according to the kepler's third law.

\begin{table}[!h]
\caption{Absolute parameters of the component stars in OW Leo}\label{absolute}
\begin{center}
\small
\begin{tabular}{lllllllll}
\hline
Parameters                        &     Primary star               &     Secondary star         \\
\hline
$M$                               & $0.31(\pm0.01)M_\odot$         & $0.95(\pm0.03)M_\odot$         \\
$R$                               & $0.63(\pm0.01)R_\odot$         & $1.04(\pm0.01)R_\odot$         \\
$L$                               & $0.43(\pm0.01)L_\odot$         & $1.01(\pm0.02)L_\odot$         \\
\hline
\end{tabular}
\end{center}
\end{table}

Although the researches on contact binaries have a very long history, the formation and evolution theory of contact binaries are still incomplete. \citet{1976ApJS...32..583W} proposed that contact binaries are evolved from initially detached binary systems. The more massive component in a binary system expanded first and filled its critical Roche lobe, and then transfer masses to the less massive companion star through the inner Lagrangian point of the binary system. In recent years, more and more contact binaries were reported to have additional components orbiting around the central binaries \citep{2019AJ....157..111Y,2019PASJ...71...39Z,2020RAA....20...47Z}. The additional components may be ejected from the center of initially multiple star system and pump out angular momentum from the multiple star system, and left out a tight binary system, which will evolve to a contact binary system finally, especially for the formation of M-type contact binaries \citep{2004A&A...414..633G,2015ApJ...798L..42Q}. Basing on the statistical research of stellar atmospheric parameters of EW-type binaries from the LAMOST, \citet{2019RAA....19...64Q} pointed out that the metal abundance of over $80\,\%$ EW-type binaries are lower than the Sun, which means that they are old population stars. They may evolve from initially detached binary systems through nuclear evolution and angular momentum loss via magnetic braking.

\begin{figure}[!ht]
\begin{center}
\includegraphics[width=13cm]{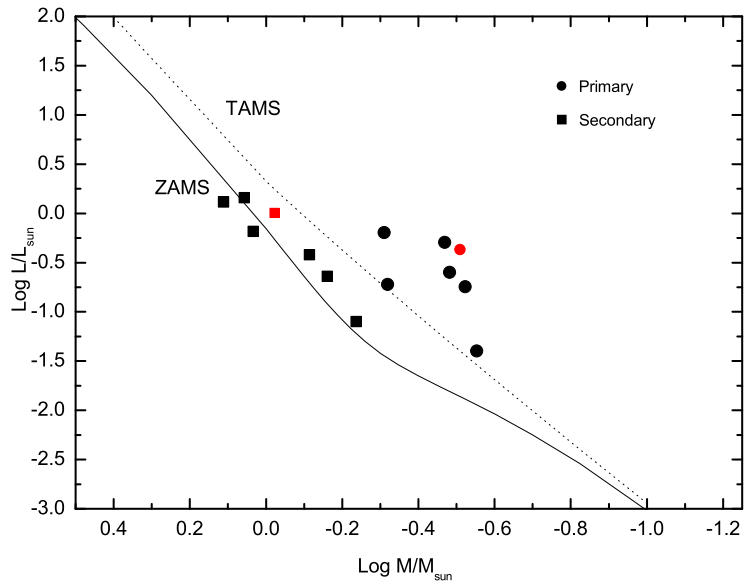}
\caption{The H - R diagram. The red circle and square represent the primary and secondary stars in OW Leo.}\label{H-R}
\end{center}
\end{figure}

According to the stellar atmospheric parameters in Table \ref{lamost}, we can see that OW Leo is a shallow contact binary with very low metal abundance. Thus, it may have not contaminate by material from an additional component. It should be evolved from an initially detached binary systems through a very long time of main sequence evolution. And the absolute parameters of several more W-subtype shallow contact binaries are collected and listed in Table \ref{shallow}. Their evolutionary status are shown in Fig. \ref{H-R}. We can see that the less massive stars have already evolved away from the terminal-age main sequence(TAMS), while most of more massive stars are still stay in the main sequence belt. For W-subtype shallow contact binaries, the more massive stars expand and fill the critical Roche lobe on its way evolving to the red giant branch (RGB), and transfer masses to the less evolved component stars. The transferred masses also filled the critical Roche lobe of the less massive star and formed the thin common envelope with the two component star embedded inside. Therefore, the less massive stars will seem more evolved than the more massive components.

\begin{table}[!h]
\caption{The W-subtype shallow contact binaries.}\label{shallow}
\begin{center}
\small
\begin{tabular}{ccllrrcc}\hline\hline
  Target                   &  $M_1$         &     $M_2$      &      $R_1$     &       $R_2$    &       $L_1$    &   $L_2$        & Ref.   \\\hline
V1007 Cas                  & $0.34M_\odot$  & $1.14M_\odot$  & $0.69R_\odot$  & $1.16R_\odot$  & $0.51L_\odot$  & $1.44L_\odot$  &    1     \\
V342 UMa 	               & $0.490M_\odot$ & $1.293M_\odot$ & $0.766R_\odot$ & $1.188R_\odot$ & $0.640L_\odot$ & $1.303L_\odot$ &    2     \\
V1197 Her 	               & $0.30M_\odot$  & $0.77M_\odot$  & $0.54R_\odot$  & $0.83R_\odot$  & $0.18L_\odot$  & $0.38L_\odot$  &    3    \\
1SWASP J104942.44+141021.5 & $0.48M_\odot$  & $0.69M_\odot$  & $0.60R_\odot$  & $0.70R_\odot$  & $0.19L_\odot$  & $0.23L_\odot$  &    4     \\
CRTS J145224.5+011522 	   & $0.28M_\odot$  & $0.58M_\odot$  & $0.44R_\odot$  & $0.61R_\odot$  & $0.04L_\odot$  & $0.08L_\odot$  &    4     \\
V2790 Ori 	               & $0.33M_\odot$  & $1.08M_\odot$  & $0.54R_\odot$  & $0.91R_\odot$  & $0.252L_\odot$ & $0.659L_\odot$ &    5     \\\hline
\end{tabular}
\end{center}
\textbf
{\footnotesize Reference:} \footnotesize (1) \citet{2018PASP..130g4201L}; (2) \citet{2019RAA....19..147L}; (3) \citet{2020RAA....20...10Z}; (4) \citet{2020AJ....159..189L};  (5) \citet{2020NewA...8001400S}
\end{table}

\begin{acknowledgements}
This research was supported by the National Natural Science Foundation of China (Grant No. 11703080, 11703082 and 11873017), the Joint Research Fund in Astronomy (Grant No. U1931101) under cooperative agreement between the National Natural Science Foundation of China and Chinese Academy of Sciences, and the Yunnan Fundamental Research Projects (Grant No. 2018FB006). We acknowledge the support of the staff of the Xinglong 85cm telescope. This work was partially supported by the Open Project Program of the Key Laboratory of Optical Astronomy, National Astronomical Observatories, Chinese Academy of Sciences. Guoshoujing Telescope (the Large Sky Area Multi-Object Fiber Spectroscopic Telescope LAMOST) is a National Major Scientific Project built by the Chinese Academy of Sciences. Funding for the project has been provided by the National Development and Reform Commission. LAMOST is operated and managed by the National Astronomical Observatories, Chinese Academy of Sciences.
\end{acknowledgements}

\label{lastpage}

\end{document}